\documentclass[12pt,english,aps,manuscript]{article}

\usepackage[]{fontenc}
\usepackage[latin9]{inputenc}
\usepackage{geometry}
\usepackage{amsbsy}
\usepackage{amssymb}
\usepackage{esint}
\usepackage[numbers]{natbib}
\usepackage{babel}
\begin{document}
\begin{center}
\textbf{\Large{}Higher Derivative Corrections to Lower Order RG Flow
Equations }
\par\end{center}{\Large \par}

\begin{center}
\vspace{0.3cm}
 
\par\end{center}

\begin{center}
{\large{}S. P. de Alwis$^{\dagger}$ }
\par\end{center}{\large \par}

\begin{center}
Physics Department, University of Colorado, \\
 Boulder, CO 80309 USA 
\par\end{center}

\begin{center}
\vspace{0.3cm}
 
\par\end{center}

\begin{center}
\textbf{Abstract} 
\par\end{center}

We show that the RG flow equation for the cosmological constant (CC)
receives contributions (in addition to those coming from the CC the
Einstein-Hilbert term and $R^{2}$ and $R_{\mu\nu}^{2}$ terms) only
from terms with just two powers of curvature, but having also powers
of the covariant derivative, in the Wilsonian effective action. In
pure gravity our argument implies that just considering $f(R)$ theories
will miss this effect which arises from terms such as $``R"\square^{n}``R",\,n=0,1,2,\ldots$.
We expect similar contributions for the flow equation of the Einstein-Hilbert
term as well. Finally we argue that the perturbative ghosts coming
from curvature squared terms in the action are in fact spurious since
they are at the cutoff scale and can be removed by (cutoff dependent)
field redefinitions.

\begin{center}
\vspace{0.3cm}
 
\par\end{center}

\vfill{}

$^{\dagger}$ dealwiss@colorado.edu

\section{Introduction}

The Wilsonian effective action \citep{Wilson:1973jj} with a cutoff
scale $\Lambda$ for a QFT (including quantum gravity with or without
coupling to matter), evaluated at some momentum or curvature scale
$\partial^{2}\phi/\phi,\,R\ll\Lambda^{2}$ may be written as an infinite
series of local operators

\begin{equation}
I_{\Lambda}=\sum\bar{g}^{A}(\Lambda)\Phi_{A}[\phi].\label{eq:ILambda}
\end{equation}
 The conceit of the asymptotic safety (AS) program is that this action
is ultra-violet complete in the sense that the limit $\Lambda\rightarrow\infty$
exists, and that only a finite (and hopefully small) number of these
operators need to be determined by experiment. In order to validate
this conjecture it is necessary to use a convenient form of the renormalization
group and a truncation of this infinite set of terms to a manageable
set. 

While there are alternative versions of the renormalization group
equation\footnote{For reviews see \citep{Morris:1998da,Bagnuls:2000ae,Rosten:2010vm}.}
for the Wilsonian action, a particularly convenient formulation that
is useful for studying the question of asymptotic safety \citep{Weinberg:1976xy,Weinberg:1979qg}\footnote{For reviews of the recent literature see \citep{Codello:2008vh,Reuter:2012id,Percacci:2017}.},
close in spirit to Polchinski's equation \citep{Polchinski:1983gv}
was derived in \citep{deAlwis:2017ysy}. In this short note we will
use this equation to argue that the RG equation for the cosmological
constant (and plausibly also the gravitational coupling constant)
does not acquire corrections from higher derivative terms such as
$R^{N},\,R_{\mu\nu}^{N},\,N>2$. In explicit calculations in a different
scheme using the Wetterich equation \citep{Wetterich:1992yh} up to
$N\sim30$ for Ricci scalar terms, these corrections have been found
to be suppressed\footnote{See for example \citep{Falls:2013bv} (and references therein) where
it is also speculated that this suppression extends to all higher
derivative terms. }. However we also point out that terms which will contribute are those
such as $R\square^{N}R,$ which (except for the case $N=0$) have
not been computed so far\footnote{Theories with higher derivatives of this form seem to have been first
considered in \citep{Shapiro:1996hz} where a calculation of the cosmological
beta function can also be found. A suggestion for performing calculations
beyond the constant curvature background calculations given in the
asymptotic safety literature (as for example in \citep{Percacci:2017})
was made recently in \citep{Steinwachs:2017bqx}.}. Finally we discuss the elimination of (perturbative) ghosts typically
associated with curvature squared terms \citep{Stelle:1976gc} by
a (scale dependent) redefinition of the metric, which will effectively
eliminate additional contributions to the propagator since all ``$R^{2}$''
terms can be removed from the action. Essentially what this shows
is that since the putative ghost is at the cutoff scale it is in fact
an artifact of the cutoff. This is of course consistent with what
happens in string theory and in fact we borrow an argument due to
Deser and Redlich \citep{Deser:1986xr} given in that context.

\section{Pure gravity case}

Let us write the Wilsonian action with a UV cutoff $\Lambda$ for
the pure gravity case as an infinite series 
\begin{eqnarray}
I_{\Lambda}^{{\rm grav}} & = & \int d^{4}x\sqrt{g}[\Lambda^{4}g_{0}(\Lambda)+\Lambda^{2}g_{1}(\Lambda)R+(g_{2a}(\Lambda)R_{\mu\nu}R^{\mu\nu}+g_{2b}(\Lambda)R^{2}+g_{3b}R_{....}R^{....})\nonumber \\
 &  & +\Lambda^{-2}[(g_{3a}(\Lambda)RR_{\mu\nu}R^{\mu\nu}+\ldots)+(g_{3a}^{(1)}R\square R+\ldots)+O(\Lambda^{-4})]\nonumber \\
 &  & +I_{\Lambda}^{({\rm G.F}.)}+I_{\Lambda}^{({\rm ghost)}}.\label{eq:action}
\end{eqnarray}
One might ask whether this series is convergent or merely an asymptotic
expansion. This question obviously arises even if the UV limit of
all the couplings of the theory exist. In the corresponding expansion
in string theory it is in fact a convergent expansion with a radius
of convergence of the order of the string scale. It is essentially
like the expansion of the propagators $1/(p^{2}+M_{s}^{2})$ in powers
of $p^{2}/M_{S}^{2}$ at low energies (with $M_{s}$ the mass of a
string excitation). For $p^{2}>M_{s}^{2}$ of course one needs to
replace the field theory by string theory. 

In the AS case one might ask what is this infinite series an expansion
of. Indeed it is not clear to the author whether this question is
even meaningful. We will nevertheless assume that the series makes
sense (converges) for $\partial^{2}/\Lambda^{2}<1$ and the UV limit
will be taken satisfying this constraint. 

The coupling constants satisfy RG equations of the form
\begin{equation}
\dot{g}^{A}+(4-n_{A})g^{A}=\eta^{A}(\{g\}),\label{eq:gdot}
\end{equation}
where $n_{A}$ is the canonical dimension of the corresponding operator
and the LHS is in principle dependent on all the couplings in the
effective action. If we truncate the system to the first two operators
then the RG equations are (with $\dot{x}\equiv\Lambda dx/d\Lambda$)
\begin{eqnarray}
\dot{g}_{0}+4g_{0} & = & \frac{1}{(4\pi)^{2}}[10e^{-g_{0}/g_{1}}-4],\label{eq:cc1}\\
\dot{g}_{1}+2g_{1} & = & -\frac{1}{(4\pi)^{2}}\frac{1}{3}[13e^{-g_{0}/g_{1}}+5].\label{eq:E-H1}
\end{eqnarray}
Now it has been observed ``experimentally'' (i.e. by explicit calculation
in truncated theories), that these equations are corrected by the
addition of $R_{....}R^{....},\,R_{..}R^{..},\,R^{2}$ terms but are
remarkably stable under the addition of higher powers of the Ricci
scalar, i.e. for a gravitational theory of the form $f(R)$.

Let us first consider the higher derivative corrections to the CC
equation since they can be computed in flat space. We will also ignore
gauge fixing and ghost action terms since they are irrelevant for
the point we wish to make. Of course this does not mean that they
do not contribute to the running of the CC of the gravitational coupling
constant, in fact their contribution has been included in (\ref{eq:cc1})(\ref{eq:E-H1}).
But here we just want to highlight the contribution of a entire class
of higher derivative operators that have been ignored hitherto. 

We will use the equation 
\begin{equation}
\Lambda\frac{d}{d\Lambda}I_{\Lambda}[\phi_{c}]=\frac{1}{2}k\frac{d}{dk}{\rm Tr}\ln K_{k,\Lambda}[\phi_{c}]|_{k=\Lambda}={\rm Tr}\exp\{-\frac{1}{\Lambda^{2}}\frac{\delta}{\delta\phi_{c}}\otimes\frac{\delta}{\delta\phi_{c}}I_{\Lambda}[\phi_{c}]\},\label{eq:RGeqn}
\end{equation}
which was derived in \citep{deAlwis:2017ysy} to calculate the contributions
of higher derivative terms in (\ref{eq:action}) to the RHS of (\ref{eq:cc1})\footnote{In the RHS of the first equality of (\ref{eq:RGeqn}) we've introduced
the kinetic operator $K$ of the theory in a background field $\phi_{c}$
with an IR cutoff $k$ and a UV cutoff $\Lambda$. So in terms of
the proper time representation we've defined ${\rm ln}K_{k,\Lambda}[\phi_{c};x,y]=-<x|\int_{1/\Lambda^{2}}^{1/k^{2}}\frac{ds}{s}e^{-\hat{K}[\phi_{c}]s}|y>$.
For more details see \citep{deAlwis:2017ysy}.}. These terms can be calculated without invoking the full machinery
of the covariant heat kernel expansion (and its generalizations).
This is because after performing the differentiation on the RHS of
(\ref{eq:RGeqn}), the curvatures can be set to zero, as what we need
on the LHS is the coefficient of the unit operator.

Let us begin with the flat space heat kernel,

\begin{equation}
H_{0}(x;y)=<x|e^{-s_{0}\hat{{\bf p}}^{2}}|y>=\frac{1}{(2\pi)^{4}}\left(\frac{\pi}{s_{0}}\right)^{2}e^{-(x-y)^{2}/s_{0}};\,\,s_{0}\equiv\frac{1}{\Lambda^{2}},\,\hat{{\bf p}}=-i\boldsymbol{\partial}\label{eq:H0}
\end{equation}
The last step was taken by going to momentum space and doing a Gaussian
integral. This simple relation is the basis of the heat kernel expansion
in curved space but its applicability is clearly restricted to two
derivative kinetic terms. However here we need to evaluate the heat
kernel when the kinetic term contains all higher powers of ${\bf \hat{p}}^{2}$
(as well as more complicated operators which we ignore here). Consider
then an operator of the form
\[
\hat{{\bf K}}={\bf \hat{p}}^{2}+s_{0}h_{1}({\bf \hat{p}}^{2})^{2}+s_{0}^{2}h_{2}({\bf \hat{p}}^{2})^{3}+\ldots={\bf \hat{p}}^{2}+\sum_{n=1}^{\infty}s_{0}^{n}h_{n}({\bf \hat{p}}^{2})^{n+1}.
\]
Now we may write 
\begin{eqnarray}
H(x;x) & = & <x|e^{-s_{0}\hat{{\bf K}}}|x>=e^{-\sum_{n=1}^{\infty}s_{0}^{n+1}h_{n}\frac{\partial^{n+1}}{\partial s^{n+1}}}<x|e^{-s\hat{{\bf p}}^{2}}|x>|_{s=s_{0}}\nonumber \\
 & = & e^{-\sum_{n=1}^{\infty}s_{0}^{n+1}h_{n}\frac{\partial^{n+1}}{\partial s^{n+1}}}\frac{1}{(2\pi)^{4}}\left(\frac{\pi}{s}\right)^{2}|_{s=s_{0}}\equiv\frac{1}{16\pi^{2}}G({\bf h})\Lambda^{4}.\label{eq:H}
\end{eqnarray}
Suppose we are interested in the contributions to the RHS of the RG
equation for the CC - namely the equation (\ref{eq:cc1}). Then in
calculating the relevant terms in $\hat{{\bf K}}$ we can set the
background fields (in particular the curvatures) to zero after performing
the differentiation that define the background field dependent kinetic
operator. Schematically this is tantamount to setting $\delta R\sim\kappa\square\delta g$
and taking the limit $R\rightarrow0$ after differentiation. Clearly
the contributions to ${\bf K}$ will only come from the Einstein-Hilbert
term, the $``R^{2}"$ terms and terms of the form $R\square^{n}R$.
In particular none of the higher than quadratic powers of a truncation
of the form $f(R)=\sum a_{n}R^{n}$, will contribute to the RG of
the CC.

Thus the RG equation for the CC is corrected from (\ref{eq:cc1})
to be of the form 

\begin{equation}
\dot{g}_{0}+4g_{0}=\frac{1}{16\pi^{2}}[10e^{-g_{0}/g_{1}}-4+e^{-\hat{\lambda}_{1}}+G({\bf h})].\label{eq:cc2}
\end{equation}

This then explains why in $f(R)$ theories the RG equation for the
CC is not affected beyond the $R^{2}$ terms. We expect a similar
result to hold for Newton's constant $G_{N}$ though in that case
the flat space argument that we gave above will need to be modified.
For instance in analogy with the above argument one expects the class
of operators of the form $k_{n}R^{2}\square^{n}R$ to contribute to
the beta function equation for $g_{1}$ modifying (\ref{eq:E-H1})
to 
\begin{equation}
\dot{g}_{1}+2g_{1}=-\frac{1}{(4\pi)^{2}}\frac{1}{3}[13e^{-g_{0}/g_{1}}+5+H({\bf k})],\label{eq:E-H2}
\end{equation}
 where $H({\bf k})$ is in principle a computable function of the
set $\{k_{n}\}$ for which however we are unable at this point to
give an explicit expression like (\ref{eq:H}).

In other words the above calculation suggests that the RG equations
(and hence the non-trivial fixed points) for the lowest order operators
in Einstein gravity with a cosmological constant, may be seriously
affected once one includes higher derivative operators of the above
classes. Of course it may turn out that the fixed point values of
the couplings $h_{n},k_{n}$ are indeed small as was the case with
couplings of other higher dimensional operators. It is nevertheless
important check this directly since \textit{a priori} one would expect
them to be $O(1)$ numbers.

Note that the issue here is not the scaling dimensions of these higher
dimensional operators near the UV fixed point. Indeed it may well
be that these are essentially given by their canonical values with
small corrections so that all these operators would then be irrelevant.
Thus their couplings would have to be set equal to their fixed point
values so as to be on the critical surface. However if these values
are of $O(1)$ then it would be difficult to establish the existence
of an RG trajectory joining the UV fixed point of the theory to the
IR fixed point where current experiments are done.

\section{Scalar field theory}

Let us now add a (scalar) matter action of the form 
\begin{equation}
I_{\Lambda}^{{\rm matter}}=\int d^{4}x\sqrt{g}[Z(\phi^{2}/\Lambda^{2})\frac{1}{2}\phi(-\square)\phi+V(\phi,\Lambda)+\xi(\phi,\Lambda)R+O(\partial^{4})],\label{eq:actionmatter}
\end{equation}
 with 
\begin{eqnarray}
V(\phi,\Lambda) & = & \frac{1}{2}\lambda_{1}(\Lambda)\Lambda^{2}\phi^{2}+\frac{1}{4!}\lambda_{2}(\Lambda)\phi^{4}+\frac{1}{6!}\lambda_{3}(\Lambda)\Lambda^{-2}\phi^{6}+\ldots,\label{eq:Vexpansion}\\
Z\left(\frac{\phi^{2}}{\Lambda^{2}}\right) & = & Z_{0}+\frac{1}{2}Z_{1}\frac{\phi^{2}}{\Lambda^{2}}+\ldots\label{eq:Zexpansion}\\
\xi\left(\phi,\Lambda)\right) & = & \frac{1}{2}\xi_{1}\phi^{2}+\frac{1}{4!}\xi_{2}\frac{\phi^{4}}{\Lambda^{2}}+\ldots\label{eq:xiexpansion}
\end{eqnarray}
The situation here is very similar to the case of pure gravity. If
we ignore operators of the form $\phi^{n}\square^{m}\phi$ with $m>1$
then the RG equation for the potential is not affected. But inclusion
of these terms can seriously affect in principle the evolution of
the couplings in the potential. Let us consider this in more detail. 

In order to highlight the issue it is enough again to simply consider
the flat space case. Consider a set of higher derivative terms of
the form ($s_{0}\equiv1/\Lambda^{2}$) 
\[
\frac{1}{2}\phi\sum_{n=2}^{\infty}z_{n-1}s_{0}^{n-1}(-\square)^{n}\phi.
\]
The usual kinetic operator $\hat{K}=\hat{{\bf p}}^{2}$ then gets
replaced by 
\[
\hat{K}=\hat{{\bf p}}^{2}+\sum_{n=1}^{\infty}z_{n}s_{0}^{n}(\hat{{\bf p}}^{2})^{n+1}.
\]
Now if we ignore the higher derivative terms in $\hat{K}$ we have
the local potential equation for a scalar field theory;
\begin{equation}
\Lambda\frac{d}{d\Lambda}V_{\Lambda}(\phi)=\frac{\Lambda^{4}}{16\pi^{2}}e^{-Z_{0}^{-1}V^{''}(\phi)/\Lambda^{2}}.\label{eq:locpot}
\end{equation}
This gives an infinite set of RG equations for the couplings $\hat{\lambda}_{i}\equiv\lambda_{i}/(Z_{0})^{i}$
of (\ref{eq:Vexpansion}) of the form
\begin{equation}
\dot{\hat{\lambda}}_{n}+(n\dot{\gamma}+(4-2n))\hat{\lambda}_{n}=(2n)!\frac{\Lambda^{4}}{16\pi^{2}}e^{-Z_{0}^{-1}V^{''}(\phi)/\Lambda^{2}}|_{\phi^{2n}},\label{eq:lRGeqns}
\end{equation}
(with $\gamma=\ln Z_{0}$) where the RHS carries an instruction to
pick the coefficient of $\phi^{2n}$ in the expansion of the exponential
in powers of $\phi^{2}$. For instance we have for the first few couplings,
\begin{eqnarray}
\dot{\hat{\lambda}}_{1}+(2+\dot{\gamma})\hat{\lambda}_{1} & = & -\frac{e^{-\hat{\lambda}_{1}}}{16\pi^{2}}\hat{\lambda_{2}}\label{eq:l1}\\
\dot{\hat{\lambda}}_{2}+(0+2\dot{\gamma})\hat{\lambda}_{2} & = & \frac{e^{-\hat{\lambda}_{1}}}{16\pi^{2}}(3\hat{\lambda}_{2}^{2}-\hat{\lambda}_{3}),\label{eq:l2}\\
\dot{\hat{\lambda}}_{3}+(-2+3\dot{\gamma})\hat{\lambda}_{3} & = & \frac{e^{-\hat{\lambda}_{1}}}{16\pi^{2}}6!(-\frac{1}{3!\times8}\hat{\lambda}_{2}^{3}-\frac{1}{2\times4!}\hat{\lambda}_{2}\hat{\lambda}_{3}-\frac{1}{6!}\lambda_{4}),\label{eq:l3}
\end{eqnarray}
etc. Thus it would appear that one is able to iteratively solve for
a (non-trivial) fixed point ($\hat{\lambda}_{n}=\lambda_{n}^{*};\,\,\dot{\hat{\lambda}}_{n}=0,\,\dot{\gamma}=0,\,\forall n$
in terms\footnote{In general we should allow for the possibility that the anomalous
dimension $\eta\equiv\frac{1}{2}\dot{\gamma}$ at a fixed point is
non-zero as in the Wilson-Fischer case, although it is zero at the
trivial fixed point . We've ignored this possibility for simplicity
since one needs to go beyond the local potential equation to discuss
this.} of $\lambda_{1}^{*}$. For instance given a value $\lambda_{1}=\lambda_{1}^{*}\ne0$
the first equation above determines $\hat{\lambda}_{2}^{*}$, the
second equation then determines $\hat{\lambda}_{3}^{*}$, the third
$\hat{\lambda}_{4}^{*}$ and so on for all the couplings in the potential,
which is therefore determined at the fixed point as an infinite series.
This fixed point however depends on $\lambda_{1}$ which is arbitrary,
and hence we appear to have a fixed line \footnote{\label{fn:The-investigation-of}}
parametrized by $\lambda_{1}$. However this is probably an artifact
of the truncation/iteration.

In fact the investigation of scalar field theory in this so-called
local potential approximation (LPA) is an old subject with conflicting
claims\footnote{See for example \citep{Halpern:1995vf}\citep{Morris:1996nx} for
the original dispute, and \citep{Gies:2000xr}\citep{Bridle:2016nsu}
for more recent work with references to the earlier literature.}. Clearly the infinite series for the potential at a non-trivial fixed
point (with $\lambda_{1}^{*}\ne0$) will be convergent only for $\hat{\phi}\equiv\phi/\Lambda<O(1)$.
If on the other hand one requires the potential $\hat{V}=V(\phi)/\Lambda^{4}=V(\hat{\phi})$
to exist for all $\hat{\phi}$ then the continuous scaling dimension
solutions implied by the above iterative analysis will not give a
true solution and the theory will remain trivial \citep{Morris:1996nx}\citep{Bridle:2016nsu}.
Also the analysis below introduces an additional layer of uncertainty
to the claims of \citep{Halpern:1995vf,Halpern:1996dh}. 

The investigation of scalar field theory in this so-called local potential
approximation (LPA) is an old subject with conflicting claims. See
for example \citep{Halpern:1995vf}\citep{Morris:1996nx} for the
original dispute, and \citep{Gies:2000xr}\citep{Bridle:2016nsu}
for more recent work with references to the earlier literature. Clearly
the infinite series for the potential at a non-trivial fixed point
(with $\lambda_{1}^{*}\ne0$) will be convergent only for $\hat{\phi}\equiv\phi/\Lambda<O(1)$.
If on the other hand one requires the potential $\hat{V}=V(\phi)/\Lambda^{4}=V(\hat{\phi})$
to exist for all $\hat{\phi}$ then the continuous scaling dimension
solutions implied by the above iterative analysis will not give a
true solution and the theory will remain trivial \citep{Morris:1996nx}\citep{Bridle:2016nsu}.
The analysis below introduces an additional layer of uncertainty to
the claims of \citep{Halpern:1995vf,Halpern:1996dh}. 

However one cannot ignore the higher derivative terms - there is no
reason \textit{a priori }to assume that dimensionless numbers $z_{n}$
are small. The local potential equation gets replaced by (see equations
(\ref{eq:RGeqn})-(\ref{eq:H})),
\begin{equation}
\Lambda\frac{d}{d\Lambda}V_{\Lambda}(\phi)=F({\bf z})\frac{\Lambda^{4}}{16\pi^{2}}e^{-Z_{0}^{-1}V^{''}(\phi)/\Lambda^{2}},\,\,F({\bf z})\frac{\Lambda^{4}}{16\pi^{2}}\equiv e^{-\sum_{n=1}^{\infty}s_{0}^{n+1}z_{n}\frac{\partial^{n+1}}{\partial s^{n+1}}}\frac{1}{(2\pi)^{4}}\left(\frac{\pi}{s}\right)^{2}|_{s=s_{0}\equiv1/\Lambda^{2}}.\label{eq:poteqn}
\end{equation}
Thus without any knowledge of the (infinite) number of higher derivative
couplings $z_{n}$, one cannot really extract any useful information
from this (exact!) equation. In particular the fixed line that appears
to exist in the iterative solution that ignored the higher derivative
couplings, may be destabilized. One cannot really evaluate the potential
at the fixed point without knowing the function $F({\bf z}^{*})$
at the unknown fixed point values of all these higher derivative couplings
- which may or may not have finite values!

\subsection{Coupling to gravity}

Now let us consider the ramifications of the above, firstly to scalar
field theory coupled to gravity, and then to the standard model coupled
to gravity. 

Once the coupling to gravity is included the evolution of scalar field
couplings gets modified. The mass term and the quartic coupling equations
(\ref{eq:l1})(\ref{eq:l2}) for instance, get replaced by 
\begin{eqnarray}
\dot{\hat{\lambda}}_{1}+\dot{\gamma}\hat{\lambda}_{1}+2\hat{\lambda}_{1} & = & -\frac{e^{-\hat{\lambda}_{1}}}{16\pi{}^{2}}F({\bf z})[\frac{\hat{\lambda}_{2}}{2}-\frac{1}{8}g_{N}\hat{\lambda}_{1}^{2}]+\frac{5}{(4\pi)^{2}}e^{2\lambda_{CC}}g_{N}\hat{\lambda}_{1},\label{eq:l1grav}\\
\frac{1}{4!}(\dot{\hat{\lambda}}_{2}+\dot{\gamma}\hat{\lambda}_{2}) & = & \frac{e^{-\hat{\lambda}_{1}}}{16\pi{}^{2}}F({\bf z})(\frac{1}{8}\hat{\lambda}_{2}^{2}-\frac{1}{4!}\hat{\lambda}_{3}+\frac{1}{3}\frac{1}{g_{1}}\hat{\lambda}_{2}\hat{\lambda}_{1})+\frac{4g_{N}e^{2\lambda_{CC}}}{(4\pi)^{2}}\frac{1}{4!}\hat{\lambda}_{2}.\label{eq:l2grav}
\end{eqnarray}
In the above we have defined $16\pi G_{N}(\Lambda)\Lambda^{2}\equiv g_{N}\equiv-1/g_{1}$
the dimensionless Newton constant. It should be emphasized that the
last term in each of the above eqns. (\ref{eq:l1grav})(\ref{eq:l2grav})
is a quantum gravity effect and in particular the contribution to
the quartic coupling in the second equation is in fact crucial to
the Shaposhnikov-Wetterich argument for the Higgs mass\citep{Shaposhnikov:2009pv}.
Of course such terms will not appear in background gravity calculations
such as those in \citep{Buchbinder:1992rb,Shapiro:2015ova}.

Similarly the gravitational equations acquire some matter contributions
so that (\ref{eq:cc1})(\ref{eq:E-H1}) are replaced by $\lambda_{CC}\equiv g_{N}g_{0}$,
\begin{eqnarray}
\dot{\lambda}_{{\rm CC}}+2\lambda_{{\rm CC}} & = & \frac{g_{N}}{16\pi{}^{2}}\Huge[(5-\frac{13}{3}\lambda_{{\rm CC}})e^{2\lambda_{{\rm CC}}}-(2+\frac{5}{3}\lambda_{{\rm CC}})+e^{-\hat{\lambda}_{1}}(\frac{1}{2}-\frac{1}{6}\lambda_{{\rm CC}}(1-6\hat{\xi}_{1}))\nonumber \\
 &  & +\frac{3G({\bf h})-2\lambda_{CC}H({\bf k})}{6}\Huge],\label{eq:cc2-1}\\
\dot{g}_{N}-2g_{N} & = & -\frac{g_{N}^{2}}{16\pi{}^{2}}\frac{1}{3}\left[13e^{2\lambda_{{\rm CC}}}+5+\frac{1}{2}e^{-\hat{\lambda}_{1}}(1-6\hat{\xi}_{1})+H({\bf k})\right].\nonumber 
\end{eqnarray}

If one ignores the higher derivative corrections $F,G,H$ that we
discussed above, this system appears to admit a non-trivial fixed
point. For instance the last two equations along with the (\ref{eq:l1grav})
seems to have a fixed point solution in terms of two undetermined
parameters $\hat{\lambda}_{1},\lambda_{{\rm CC}}$ which can be treated
as free parameters.

However once one includes the above mentioned higher derivative operators
these equations are corrected by the function $G({\bf h})$ defined
in (\ref{eq:H}), another function $H({\bf k})$ of the couplings
${\bf k}$ of the $R\square^{n}R^{2}$ type terms (see equation (\ref{eq:E-H2})),
and the function $F({\bf z})$ in (\ref{eq:l2grav}). Unfortunately
\textit{a priori }there does not seem to be any reason why these extra
terms should be negligible at a fixed point. In particular the existence
of a (physically required) positive fixed point value $g_{N}^{*}$
would appear to depend on the sign and magnitude of $H({\bf k}^{*})$.
However it is perhaps reasonable to assume that the infinite series
represented by $H(k^{*})$ converges to an $O(1)$ number leaving
the original conclusion about $g_{N}^{*}$ unchanged.

\subsection{The standard model}

One of the striking successes of the asymptotic safety program is
the calculation of the Higgs mass \citep{Shaposhnikov:2009pv}. This
was in fact a prediction, since the calculation was done well before
the Higgs was discovered at the LHC. So it is important to understand
how this came about and what assumptions went into it. In particular
one may worry that the series of higher derivative terms discussed
above may affect this calculation and destroy its agreement with experiment.

However the prediction of the Higgs mass is actually unaffected by
these extra terms provided the assumption of a non-trivial fixed point
for the standard model coupled to gravity remains valid (with positive
$g_{N}^{*}$), in the presence of these extra terms, and the sign
of the gravitational correction to the beta function of the Higgs
self-coupling remains unchanged. The latter is essentially the last
term on the RHS of (\ref{eq:l2grav}) and appears to be independent
of the additional corrections discussed in this note. In particular
of course the existence of a trajectory connecting the trivial IR
fixed point to the non-trivial UV fixed should persist  in the presence
of these terms. If that is the case the crucial argument \citep{Shaposhnikov:2009pv}
that the Higgs self-coupling is irrelevant at the UV fixed point,
so that its value should be set to its fixed point value namely zero,
will persist and we would recover the prediction of the Higgs mass.
The same appears to be true for the calculation of the top quark mass
in \citep{Eichhorn:2017ylw}\footnote{In \citep{Eichhorn:2017eht} it has been argued that higher dimension
operators may not affect the AS results for standard model couplings.}.

\section{Elimination of curvature squared terms and removing a ghost}

Finally we would like to make some remarks about the curvature squared
terms that we have considered in this paper and the issue of the spin
two ghost that appears to be present in such theories \citep{Stelle:1976gc}. 

It is well known that curvature squared terms $R^{2},R_{\mu\nu}^{2},R_{\mu\nu\lambda\sigma}^{2}$
terms can be replaced by the Euler density - which gives no contribution
to the scattering amplitudes. This is accomplished by a (scale dependent
in our case) (metric) field redefinition of the form 
\begin{equation}
g_{\mu\nu}\rightarrow g_{\mu\nu}+b_{0}R_{\mu\nu}+b_{1}Rg_{\mu\nu}\label{eq:gtrans}
\end{equation}
 which (with an appropriate choice of the coefficients $b_{i}$) will
remove the Ricci scalar and tensor squared terms in equation (\ref{eq:action})
and replace the Riemann squared term by the Euler density. As is well
known such a redefinition is expected to leave the physical results
of the theory (the S-matrix) unchanged. Of course this does not mean
that the RG equations for the coefficients $g_{2a},g_{2b},g_{2c}$
are eliminated since in deriving these equations we worked in a field
basis which is cutoff independent - the cutoff dependence coming entirely
from the coupling constants. However in order to accomplish the replacement
of the quadratic terms by the Euler density, the field redefinition
above clearly must be cutoff dependent - thus we can no longer take
the redefined fields to be cutoff independent, hence the flow of these
terms will contribute in a different way. Also this field redefinition
will change all higher order couplings in a $b_{i}$ dependent way,
though since the original action by definition must contain all terms
which are consistent with the symmetries of the theory this will simply
redefine the coefficients of these terms.

What is however somewhat less well known is the fact that all terms
that are quadratic in curvature with additional derivatives can be
eliminated by a (scale dependent) field redefinition. This appears
to have been first observed by Deser and Redlich \citep{Deser:1986xr}.
They argued that by making repeated use of Bianchi identities any
term in the action involving just two Riemann tensors, but also (necessarily
even say $=2n\ne0$) powers of the covariant derivative, can always
be written in the form 
\[
4R_{\mu\nu}\square^{n}R^{\mu\nu}-R\square^{n}R+O(R^{3})+{\rm total\,derivatives}.
\]
Thus the entire collection of terms in the action that are just quadratic
in the curvatures (but with arbitrary numbers of derivatives), up
to the Euler density term, total derivatives and higher powers of
the curvature can be written in the form 
\begin{equation}
{\cal L}\sim\sqrt{g}[a_{1}R_{\mu\nu}R^{\mu\nu}+a_{2}R^{2}+\sum_{n=1}(a_{1}^{(n)}R_{\mu\nu}\square^{n}R^{\mu\nu}+a_{2}^{(n)}R\square^{n}R)].\label{eq:R2terms}
\end{equation}
However all such terms can be removed by a generalization of the field
redefinition (\ref{eq:gtrans}) of the form,
\begin{equation}
g_{\mu\nu}\rightarrow g_{\mu\nu}+b_{0}R_{\mu\nu}+b_{1}Rg_{\mu\nu}+\sum_{n=1}(b_{1}^{(n)}\square^{n}R_{\mu\nu}+b_{2}^{(n)}\square^{n}R).\label{eq:gtrans1}
\end{equation}
Thus one can eliminate all curvature squared terms (apart from the
Euler density which does not give any contribution to amplitudes in
perturbation theory), from the Lagrangian. This implies that the graviton
propagator in this transformed basis is the same as in the Einstein
theory. Hence it appears that one could eliminate the perturbative
spin two ghost found in $R^{2}$ theories by Stelle \citep{Stelle:1976gc}.
Similar arguments could be used to eliminate all higher derivative
terms which are quadratic in the scalar field $\phi$ such as $\phi\square^{n}\phi$.

Let us now discuss the relevance of this observation to the discussion
of higher derivative contributions to the beta functions in this paper,
and more generally to the AS program as a whole.

The coefficients $a_{i}^{(n)},\,n\ge0$ in equation (\ref{eq:R2terms})
are cutoff dependent. Therefore the coefficients $b_{1}^{(n)}$ are
necessarily cutoff dependent. However as mentioned earlier, when we
computed the beta functions we worked in a basis where the fields
(including the metric), were taken to be independent of the cutoff
with the only dependence on $\Lambda$ coming from the coupling constants.
If we assumed that for the old basis of fields, this is no longer
true for the new basis so one would have to rederive the beta function
equations, with the metric now being dependent on the cutoff. Alternatively
one may start with the basis in which the $R^{2}$ terms are absent,
and take the fields to be independent of the cutoff as before. However
the RG evolution will generate these terms again. Clearly this is
no different from saying that even if one took as an initial condition
at some arbitrary scale the coefficients $a_{i}^{(n)},\,n\ge0$ to
be zero, the RG flow will generate them. Thus it is clear that the
additional terms in the beta function equation for the cosmological
constant (\ref{eq:cc2})(\ref{eq:cc2-1}) will need to be taken into
account. It is also clear from the above that similar arguments can
be made for fields (such as the scalar field treated above), to remove
the apparent ghosts that result from higher derivative terms that
are quadratic in the fields and which are generated by RG flows even
if set to zero at any given scale.

On the other hand the Deser-Redlich argument implies that at any arbitrary
scale at which we do perturbation theory, one can always find a basis
in which these ghosts \citep{Stelle:1976gc} are absent. Hence in
a theory which incorporates all higher derivative terms allowed by
symmetries such ghosts are spurious. This means that this particular
ghost problem is not an issue for the AS program. Similarly it is
clear that the scalar ghost(s) coming from higher derivative terms
that are quadratic in $\phi$ are also spurious. Clearly this is a
consequence of the fact that these ghosts appear at the cutoff scale,
which if the AS conjecture is valid can be pushed to arbitrarily high
scales.

It is important to point out that the problem identified in \citep{Stelle:1976gc}
relates to a theory with just the Einstein term and the curvature
squared terms with no higher derivative terms. The reason the latter
can be set to zero is that this theory is renormalizable. On the other
hand if we make the field redefinition (\ref{eq:gtrans}) to get rid
of these terms to eliminate the additional (quartic) terms in the
inverse propagator, we necessarily generate higher powers of curvature.
This is because the coefficients of additional terms are not infinitesimal,
so we are really doing a finite shift of the field variable resulting
in an infinite functional Taylor series. Thus the price of removing
the ghost is the loss of renormalizabilty and hence predictivity at
high (close to Planck) scales - unless of course the theory is asymptotically
safe!

A different way of making the same point is that in $"R^{2}"$ gravity
theories one has a fixed finite scale - the Planck scale. The aim
is then to calculate scattering amplitudes not just at low energies
but also at arbitrary energies $E>M_{P}$. Since the theory is renormalizable
this is possible but given that there is a spin two ghost at $E=M_{P}$
the theory cannot be unitary. In the AS case the situation is radically
different. Here the calculations that can be performed are always
at energies $E<\Lambda$ which is an arbitrary cutoff and if AS is
valid can be taken to be arbitrarily large. The ghost in this formulation
is always at the scale $\Lambda$ and hence is beyond the regime of
validity of the calculation - hence it is spurious. 

When one has an infinite number of higher derivative terms, as will
be the case when these ghosts are eliminated, and in any case is a
given in the AS program, one is immediately confronted with questions
of convergence and non-locality. In fact the original motivation of
the Deser-Redlich paper was to understand how the low energy expansion
of string theory avoids having perturbative ghosts as it must, since
string theory (at least around asymptotically flat backgrounds) is
free of ghosts. In string theory this infinite series is expected
to sum up to a non-local string field theory. Unfortunately (in the
closed string - i.e. gravity case) there is no closed form action
(unlike for the open string). What we do have are expressions for
the scattering amplitudes (involving gravitons) in a flat background\footnote{The fact that the simplest background solution for string theory is
flat 10 (or 26) dimensional space is being ignored in this discussion.
Effectively what is being assumed is that the space has been compactified
in to a 4d space with some six dimensional space with fixed moduli
and or some abstract CFT with the appropriate central charge.}, giving well defined unitary analytic scattering amplitudes (in principle)
to all orders in perturbation theory. It is unclear what the meaning
of the corresponding infinite series in the AS program is (in other
words the interpretation of the sum), if the AS program is an alternative
to, rather than a complementary way of looking at, string theory. 

On the other hand one may take a more pragmatic view of the whole
program. In fact this appears to be the point of view taken by Weinberg
(see for example \citep{Weinberg:2009wa}). In this approach suppose
one wishes to calculate (say) the amplitude for a scattering process
involving gravitons at some energy scale $E$. Then the cutoff $\Lambda$
should be chosen such that $\Lambda>E$ but not too much larger, so
that the calculation to some low order in perturbation theory will
suffice, assuming of course that the AS program is valid, and that
the (finite number of) relevant couplings at this scale have been
determined by the RG and the (infinite number of) irrelevant couplings
are all determined by their fixed point values. Now the above discussion
of ghost elimination means that we still need to find the field redefinition
(at this value of the cutoff), that eliminates all the higher derivative
propagator terms. In any practical calculation presumably one needs
only to determine a finite number of such terms (how many will depend
on the desired accuracy of the calculation), since all but a finite
set can be ignored. Of course given that the S-matrix should be independent
of field redefinitions all that the above argument establishes is
that an apparent (perturbative) ghost at the cutoff scale is really
a spurious state. Thus as long as one is working at an energy scale
which is below this cutoff the scattering amplitudes should not be
affected.

It should be stressed that the above procedure of ghost elimination
requires (for arbitrarly high cutoff scale) a quasi-local field redefinition.
This is to be expected since the starting point itself is quasi local
consisting of an infinite series in the derivative expansion. One
expects that such quasi local field redefinitions (unlike for instance
the strictly non-local field redefinitons such as that which leads
to the Nicolai transformation), would leave the S-matrix unchanged.
Finally let us stress that the elimination of the higher derivative
quadratic in curvatures/fields terms which pertain to putative ghosts,
depends on an iterative procedure that is valid to arbitrarily high
values of $\Lambda$ the cutoff, only under the assumption that the
the large cutoff limit exists - i.e. there exists an UV fixed point.
This of course is the fundamental assumption behind the AS program
and while there is considerable evidence for this, it is far from
being proven yet.

\section{Acknowledgements}

I wish to thank Astrid Eichhorn, Jan Pawlowski and Roberto Percacci
for discussions and comments on the manuscript and Stanley Deser for
comments on the last section. I also wish to thank Tim Morris for
an e-mail on the controversy related to the question of a fixed point
in scalar field theory. Finally I would like to acknowledge the hospitality
of the Abdus Salam ICTP, Trieste, Italy, where some of this work was
done and to thank the Dean of the College of Arts and Sciences at
the University of Colorado for partial support of this research.

\bibliographystyle{apsrev}
\bibliography{myrefs}

\end{document}